# Social Media Analysis for Product Safety using Text Mining and Sentiment Analysis

Haruna Isah, Paul Trundle, Daniel Neagu
Artificial Intelligence Research (AIRe) Group
School of Electrical Engineering and Computer Science
University of Bradford
Bradford, UK
H.Isah@student.bradford.ac.uk, P.R.Trundle@bradford.ac.uk, D.Neagu@bradford.ac.uk

*Abstract*— The growing incidents of counterfeiting and associated economic and health consequences necessitate the development of active surveillance systems capable of producing timely and reliable information for all stake holders in the anti-counterfeiting fight. User generated content from social media platforms can provide early clues about product allergies, adverse events and product counterfeiting. This paper reports a work in progresswith contributions including: the development of a framework for gathering and analyzing the views and experiences of users of drug and cosmetic products using machine learning, text mining and sentiment analysis; the application of the proposed framework on Facebook comments and data from Twitter for brand analysis, and the description of how to develop a product safety lexicon and training data for modeling a machine learning classifier for drug and cosmetic product sentiment prediction. The initial brand and product comparison results signify the usefulness of text mining and sentiment analysis on social media data while the use of machine learning classifier for predicting the sentiment orientation provides a useful tool for users, product manufacturers, regulatory and enforcement agencies to monitor brand or product sentiment trends in order to act in the event of sudden or significant rise in negative sentiment.

*Keywords— text mining; sentiment analysis; product safety; social media; machine learning*

## I. INTRODUCTION

The global scourge of counterfeit drug and cosmetic products poses a great threat to public safety. One strategy for combating counterfeit products is through the effective communication and tracking of early warning signals of product allergies, side or adverse effects, drug resistance and disease outbreaks [1]. Social media is now a platform for sharing virtually all kinds of information; studies on Twitter data have demonstrated that aggregating millions of messages can provide valuable insights into a population [2], hence, pharmaceutical manufacturers need to track patients' opinions on their products for decision making [3].

To address the menace of drug and cosmetic products counterfeiting, a surveillance system capable of harnessing and tracking online reported views and experiences of users of these products is needed. We propose novel research questions that include the following:

1. What is the public sentiment about a given brand (s) of drug and cosmetic product?

2. By monitoring conversation about a given brand (s) of drug and cosmetic products over a sample user population, can we be able to predict evolving adverse effects and the possibility of product counterfeiting?

We approach these problems by applying text mining and sentiment analysis techniques on drug and cosmetic product related text data. The following are the contributions of this work:

1. We propose a product safety framework for harnessing and processing the views and experiences of users of popular brands of drug and cosmetic products reported on social media platforms using text mining and sentiment analysis.

2. We utilised the framework to gather and analyse views and experiences of users of cosmetic and drug products in form of tweets and Facebook comments using lexicon and machine learning approaches.

3. We demonstrated how to develop custom lexicon and training data and also modeled a Naive Bayes classifier for sentiment prediction of views and experiences of users of popular brands of drug and cosmetic products.

This work is organised thus: Section II provides a review of related work; Section III details the structure and flow of the proposed framework; Section IV demonstrates the application of the framework using two case studies. The final section describes conclusions and future work.

## II. RELATED WORK

Text mining is a specialised domain that applies data mining techniques over text; some of the early attempts of exploratory data analysis over text are recorded in [4]. Sentiment analysis aims to identify and extract opinions, moods and attitudes of individuals and communities; the authors of [5] provided a technical survey and early work on sentiment analysis. When text mining and sentiment analysis techniques are combined in a project on social media data, the result is often a powerful descriptive or predictive tool; in [6], text mining was successful applied to extract Facebook posts for sentiment classification during the Arab Spring event.

Recent applications of sentiment analysis relevant to our study include crime surveillance; in [*7*], social media monitoring systems were reviewed and five essential elements that every monitoring system should include were suggested, while in [*8*], a framework was proposed to study the reactions, sentiments, and communication of civilians in response to terrorist attacks. The authors of [*9*] proposed two computational methods for estimating social media sentiment and reported better performance in comparison to standard techniques.

Sentiment analysis of social media data has also been applied for tracking disease outbreaks; the authors of [*10*] described a method for extracting tweets for early warning and outbreak detection during a Swine Flu pandemic demonstrating a strong contribution for alerting relevant stake holders for prompt action.

Twitter data allows for geographical and spatial analysis and in [*11*] a framework was developed for visualising public sentiment variations using tweets gathered based on counties and countries in the whole of UK during the event that took place following the birth of prince George in 2013; this application is particularly useful in tracking the upwards or downwards trend of product allergies in a given region over a given period of time.

The authors of [*12*] applied text mining techniques to investigate consumer attitudes towards global brands, they reported that Twitter can be used as a reliable method in analyzing attitudes towards global brands.

Pharmacovigilance is of particular interest to us, it involves the monitoring of adverse effects of pharmaceutical products. As reported in [*13*], Internet users can provide early clues about adverse drug events through their Internet surfing log data. A similar work is found in [*14*], where the visualization of CHFpatients.com forum chat sentiments was used to measure the effectiveness of a drug through quantifying its side effects, particularly for the benefit of the forum members and their physicians. We aim to apply the same approach to drug and cosmetic product users, with the intended beneficiaries been the product users, manufacturers, regulatory and enforcement agencies.

Sentiment analysis can be approached as one or combination of supervised, semi-supervised and unsupervised classification tasks; lexicon and machine learning are the popular approaches; the lexicon based approach as detailed in [*5*], [*15*] and [*16*] uses dictionaries of words annotated with their semantic orientations; the learning based approach also described in [*5*] and [*17*] require creating a model by training a classifier with labeled examples; and finally the utilization of the combination of both approaches as described in [*18*]. A recent review on these techniques was presented in [*17*]; two performance issues discovered with regard lexicon approaches are: 1) how to deal with context dependent words and 2) how to address multiple entities with varying orientations within a single sentence; one of the suggested approach for tackling these issues is the use of holistic lexicon described in [*15*] which involve exploiting external evidences and linguistic conventions of natural language expressions. The machine learning approach is reported to outperform the lexicon approach, yet suffers a general drawback of labeling large training data.

Three machine learning based classifiers, Naive-Bayes, Maximum Entropy and Support Vector Machines and a hybrid technique, label propagation, were investigated in [*17*] and a number of issues that need to be addressed before using these techniques for sentiment classification were outlined. The reported drawback of the Naive-Bayes classifier is the assumption that features are independent of each other; Maximum Entropy suffers from over-fitting in the event of sparse data. It is, however, reported that its performance can be improved by introducing a priori for each feature; Support Vector Machines was reported to outperform the other techniques, its major drawback however is the difficulty in identifying the important words that influenced the classification process due to its black box nature. The review outlined a semi-supervised method called label propagation; the technique improves the accuracy of the classification process by using the Twitter follow graph.

Driven by these motivations, especially the successes of text mining and sentiment analysis on social media data for surveillance [*7*], [*8*], [*9*] and [*10*], and brand reputation monitoring [*12*] applications and also in sentiment classification framework design in [*8*], [*11*] and [*17*], we proposed a novel framework for harnessing and processing the views and experiences of consumers of popular brands of drug and cosmetic products reported as status updates, tweets or comments on social media platforms. We also aim to adapt a combination of these approaches and computational intelligent techniques such as fuzzy sentiment scoring, in the framework for performance comparison with the traditional lexicon and other methods. A detailed description of the proposed framework is provided in the next section.

### III. THE PROPOSED FRAMEWORK

The architecture of the proposed framework for harnessing and processing the views and experiences of customers of popular brands of drug and cosmetic products is presented in Fig. 3.1

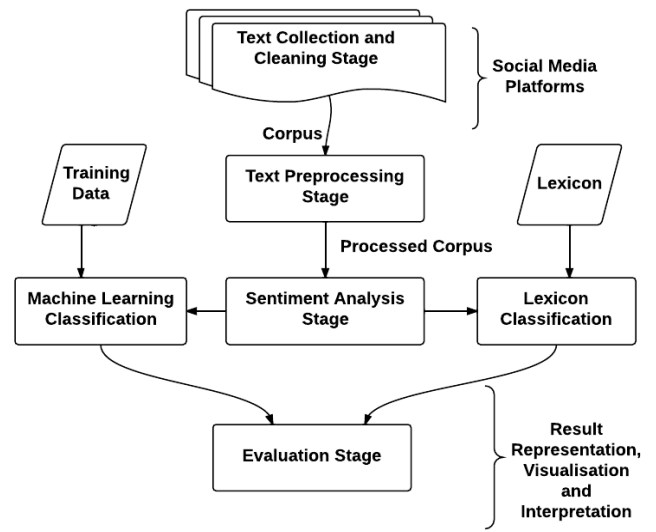

Fig. 3.1. Architecture of the proposed framework

The framework comprises four stages: text collection and cleaning; pre-processing; sentiment analysis and finally evaluation. The following is a brief description of these components:

*A. Text Collection and Cleaning Stage*

Most organisations and businesses including social media platforms now create Application Programming Interfaces (APIs) to share data. At the text collection and cleaning stage, an API call for authentication and data extraction is invoked on Facebook Graph and Twitter APIs. The framework is designed to accommodate virtually all available social media APIs, but dur to the dynamic nature of these APIs and for the purpose of this report, greater focus will be given to Twitter and Facebook. The Twitter API consists of the REpresentational State Transfer (REST) and Streaming APIs [19]. The REST API provides methods for authenticating applications, processing requests, handling imposed limits, etc. The Streaming API provides client applications with Twitter's global stream (public, user and site) of data. The Facebook Graph API provides the means of getting data into and out of the Facebook social graph. The framework employs both the REST and Streaming APIs for searching and fetching tweets, while the Facebook Graph API is used for fetching pages, status updates and comments suggesting user experiences and views on drugs and cosmetic products. Fig. 3.2 describes the data collection workflow.

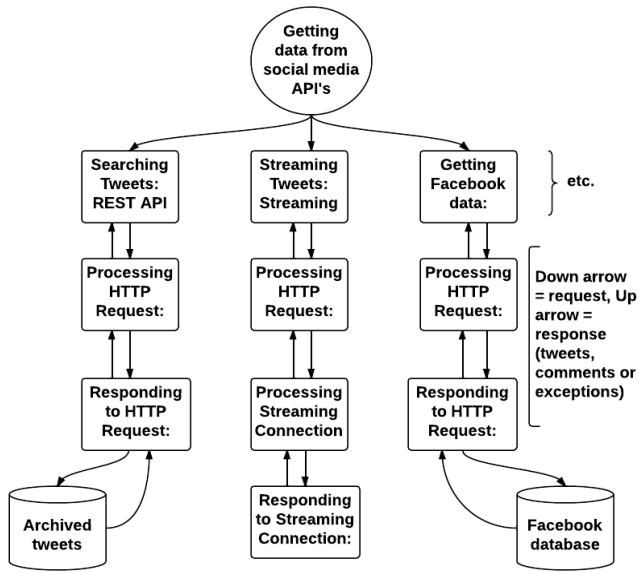

Fig. 3.2. Data Collection workflow

The collected text is noisy and in JavaScript Object Notation (JSON) formatand methods for cleaning and parsing of the data to form a corpus, i.e. a collection of comments and tweets, are incorporated for further processing.

*B. Pre-processing Stage*

At this stage, the corpus is transformed into feature vectors; for our purpose of conducting this work, we adapted the bag of words representation described in [20] and [21] due to its simplicity and because preserving the order of the features in the corpus is not of particular interest in the application. We adapted a simple feature selection or pre-processing method to transform or tokenize the text stream to words; these methods constitute a sequence of the following tasks; removing delimiters, converting all words to lower case, removing numbers and stop words [21], stemming words to their base and some application or domain specific feature transformations. The tokens are then represented as bag-of-words sparse matrix using the term frequency-inverse document frequency ($tf - idf$) weighing scheme described in [20] and [21].

We define our corpus as $C$, containing N documents defined as $d_i$, where $i = 1 \ldots N$, and tweets tokenized as words or terms $t$. The $tf - idf$ weighting scheme takes into account the relative importance of the word in the document and assigns to term $t_j$ a weight in document $d_i$ given by:

$$tf - idf(t_j, d_i) = tf(t_j, d_i) * idf(t_j) \qquad (1)$$

where:

$tf(t_j, d_i)$ denotes term frequency, the number of word occurrences in a document;

$idf(t_j) = log2\left(\frac{N}{df t_j}\right)$ denotes inverse document frequency, with $df(t_j)$ representing the number of documents containing the word. Fig 3.3 describes the pre-processing workflow.

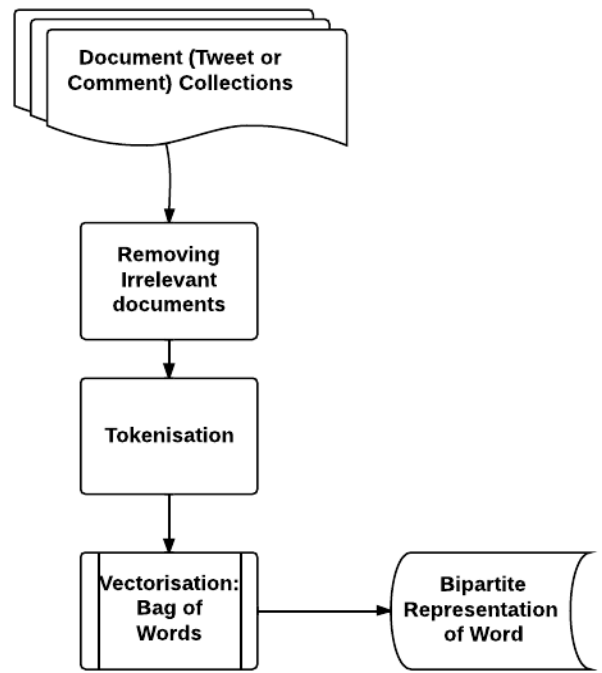

Fig. 3.3. Preprocessing workflow

Further sparseness is handled by selecting terms that appear in a minimum number of documents. The resulting bipartite representation is the input on which further tasks are performed.

## C. Sentiment Analysis Stage

This stage handles the polarity measurement, sentiment classification and clusterisation of the entire corpus and for some targeted entities. We approach these tasks by employing both lexicon and learning methods.

### 1. Lexicon-based sentiment classification

In the lexicon based approach, beside the corpus, a fundamental requirement is a pre-labelled word list or polarity lexicon. For an improved classification result as described in [22], the framework merges two lexicons, application or domain specific preassembled lexicon and a generic English based lexicon developed and being maintained by the authors of [23]. As described in Fig. 3.4, another requirement for the lexicon based classifier is a sentiment scoring function, of which there are several options, one of the most basic polarity computational schemes is described in [24]; all the words in the corpus or target collection are compared to the words in the lexicon, the overall sentiment score of the corpus or a subset will then be the difference between the numbers of positively and negatively assigned words. Therefore, the associated polarity score for each comment or tweet in the corpus is given by:

$$Score = \sum_i^n pw - \sum_j^m nw \qquad (2)$$

where $pw$ and $nw$ denote positive and negative words respectively;

A comment or tweet has an overall positive sentiment if Score > 0,

A comment or tweet has an overall neutral sentiment if Score = 0,

A comment or tweet has an overall negative sentiment if Score < 0,

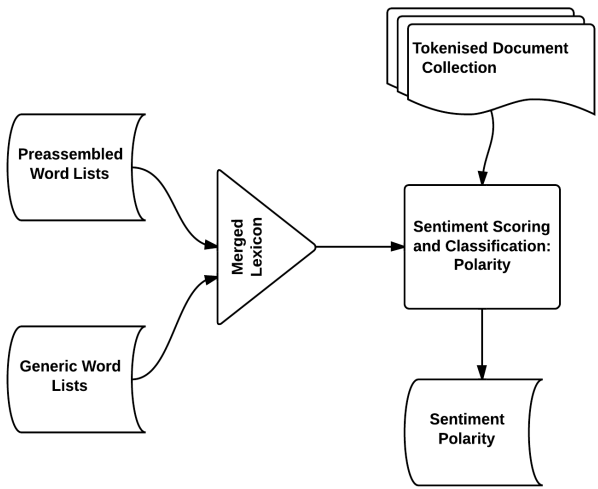

Fig. 3.4. Lexicon based sentiment classification workflow

The total score for the corpus is visualized and evaluated with simple descriptive statistics i.e. histogram and box plot. A more advanced scoring scheme include fuzzy reasoning, which is a computational intelligence technique that can be applied to improve the text classification and clustering tasks; the technique was used in [25] to generate intuitive fuzzy numbers for 150 words formed from a feature word list for sentiment classification of hotel management reviews, with higher accuracy and recalling rate.

### 2. Machine learning-based sentiment classification

In the machine learning based approach, beside the corpus, the fundamental requirement is a training dataset, already coded with sentiment classes. As described in Fig. 3.5, the classifier is trained or modeled with the labeled data such that new but similar documents are tested with the resulting model to have it predict the direction of the sentiment of the new documents. For the purpose of this report, Naive Bayes is used as a baseline classifier because of its efficiency as reported in [21]. We assume the feature words are independent and then use each occurrence to classify tweets or comments into its appropriate sentiment class; this is called multinomial event model.

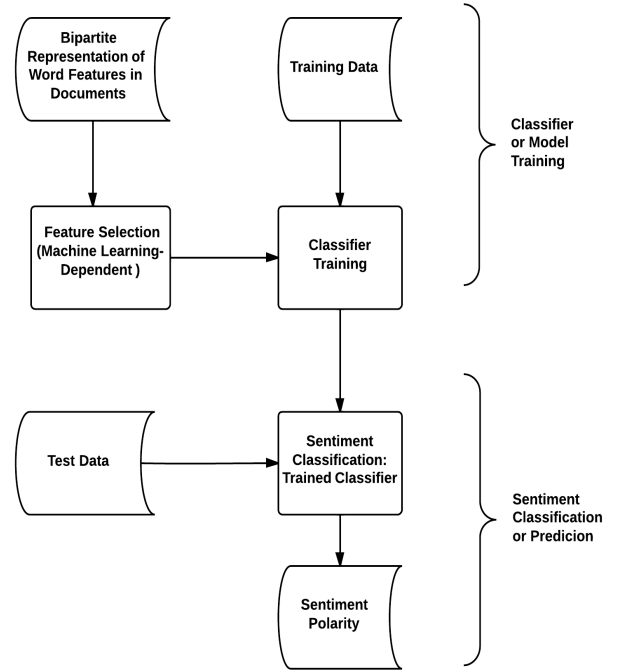

Fig. 3.5. Machine learning sentiment classification workflow

It follows from [21] that our classifier which utilises the maximum a posteriori decision rule can be represented as:

$$c_{map} = \underset{c \in C}{\operatorname{argmax}} \big(P(c|d)\big)$$
$$= \underset{c \in C}{\operatorname{argmax}} \left( P(c) \prod_{1 \leq k \leq n_d} P(t_k|c) \right) \qquad (3)$$

where $t_k$ denotes the words in each tweet or comment and $C$ the set of classes used in the classification; again, $P(c|d)$ is the conditional probability of class $c$ given document $d$, $P(c)$ the prior probability of class $c$ and the $P(t_k|c)$ conditional probability of word $t_k$ given class $c$. To estimate the prior parameters, equation (3) is then reduced to:

$$c_{map} = \underset{c \in C}{\mathrm{argmax}} \left( \log P(c) + \sum_{1 \leq k \leq n_d} \log P(t_k|c) \right) \quad (4)$$

To handle zero probabilities that may arise when a word does not occur in a particular class, $(tf - idf)$ weighing or Laplace smoothing by adding 1 to each count is employed; with Laplace smoothing, $P(t|c)$ becomes:

$$P(t|c) = \frac{T_{ct}+1}{\sum_{t' \in V}(T_{ct'}+1)} + \frac{T_{ct}+1}{(\sum_{t' \in V} T_{ct'})+B'} \quad (5)$$

where B' refers to the number of terms contained in the vocabulary V.

*D. Evaluation Stage*

The lexicon based sentiment classification result is represented as a histogram of polarity measures and the result is evaluated with reference to a ground truth or by human judgment. We use contingency tables or truth table to represent the output of the classifier and a baseline result for performance comparison.

## IV. CASE STUDIES

*A. Text Mining and Sentiment Analysis of Facebook Comments*

To demonstrate the usefulness of the proposed framework, we collected user comments and opinions from Facebook pages of 3 popular brands of drug and cosmetic related products (Avon, Dove and OralB) using the Facebook Graph API. For privacy issues, the brand names will be randomly coded as Brand X, Brand Y and Brand Z. We initially retrieved the public contents of the targeted pages and then extracted user comments and opinions from popular posts suggesting product advertisement that: 1) do not offer a prize in return i.e. Brand Y and Z and 2) offer a prize in return i.e. Brand X. TABLE 4.1 is a summary of the collected comments.

TABLE 4.1. SUMMARY OF COLLECTED COMMENTS

| Brand Name | Total number of posts retrieved | Total number of comments extracted |
|---|---|---|
| Brand X | 5000 | 654 |
| Brand Y | 665 | 1747 |
| Brand Z | 5000 | 957 |

Over the entire corpus, we apply tokenization, stop word removal and conversion to lower case functions to obtain only individual word features. The features were then represented as bag of word model with the $(tf - idf)$ weighing scheme to generate a sparse matrix, limiting the output to a minimum of three characters word length. The sparse matrix was handled using a function that remove sparse terms which have at least a 99 percentage of sparse elements. We then perform frequent term analysis to generate popular words for creating our preassembled lexicon, we adapted the method used in [22] by using the English dictionary and Thesaurus. The preassembled lexicon is then merged with social media slangs and generic lexicon for sentiment classification. A comparison lexicon sentiment analysis was performed over the 3 different brands. The result is presented in Fig 4.1. It is interesting to see the overall sentiment on all the three brands been positively skewed, with negative:neutral:positive ratios for Brand X, Brand Y and Brand Z approximately 1:42:175, 1:3:3 and 1:5:5 respectively; the high positive sentiment on Brand X confirms comments been prize driven.

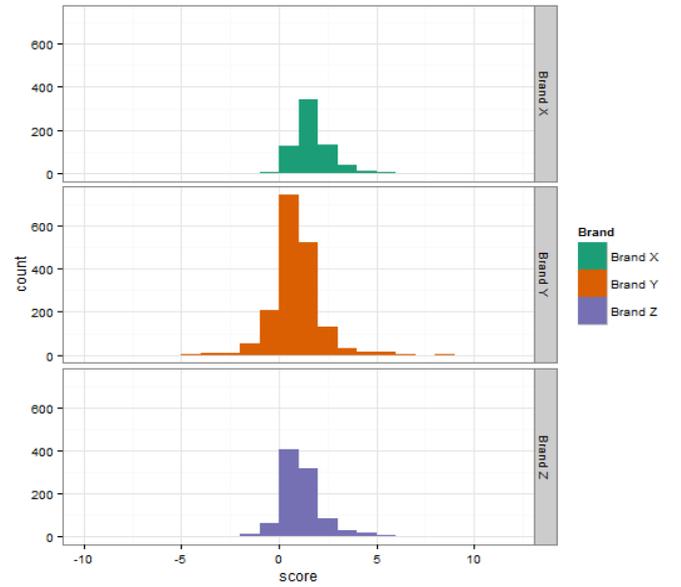

Fig. 4.1. Comparison sentiment analysis for Brand X, Brand Y and Brand Z

The distribution of these scores is presented in TABLE 4.2.

TABLE 4.2. DISTRIBUTION OF SENTIMENT SCORES FOR THE 3 BRANDS

| Brand | Sentiment Scores | | | |
|---|---|---|---|---|
| | Negative | Neutral | Positive | Row Total |
| Brand X | 3 | 127 | 524 | 654 |
| Brand Y | 282 | 742 | 723 | 1747 |
| Brand Z | 85 | 408 | 464 | 957 |
| Column Total | 370 | 1277 | 1711 | 3358 |

We also extracted three separate datasets in Brand Y each with the mention of the three generic products, soap, cream and deodorants and then performed comparison sentiment analysis on the aggregated data; the distribution of these scores is presented in TABLE 4.3. This is another interesting result with Soap been more positively skewed among the three products; the negative:neutral:positive ratio of Cream, Deodorant and Soap approximately been 1:2:2, 1:1:2 and 1:2:5 respectively.

TABLE 4.3. DISTRIBUTION OF SENTIMENT SCORES FOR THE 3 PRODUCTS

| Product | Sentiment Scores | | | Row Total |
|---|---|---|---|---|
| | Negative | Neutral | Positive | |
| Cream | 5 | 11 | 11 | 27 |
| Deodorant | 21 | 23 | 46 | 90 |
| Soap | 11 | 26 | 56 | 93 |
| Column Total | 37 | 60 | 113 | 210 |

The result is presented in Fig. 4.2.

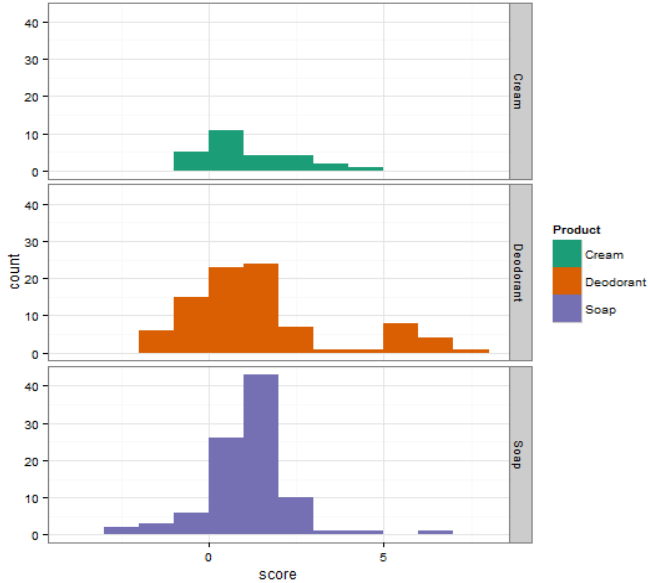

Fig. 4.2. Comparison sentiment analysis for soap, cream and deodorant

We illustrate the same problem of classifying the sentiment orientation of the entire corpus with a naïve Bayes sentiment classifier described in [26], modeled with polarity and emotional lexicon. The method classify_polarity classifies the comments as positive, neutral or negative. The classification results over the entire corpus for both methods are compared in TABLE 4.4.

TABLE 4.4. COMPARISON OF LEXICON AND MACHINE LEARNING SENTIMENT SCORES OVER THE ENTIRE COMMENT CORPUS

| Method | Sentiment Scores | | | Total |
|---|---|---|---|---|
| | Negative | Neutral | Positive | |
| Lexicon | 539 | 1436 | 1383 | 3358 |
| Naïve Bayes | 554 | 368 | 2436 | 3358 |

The negative scores for both methods agree closely while there is a sharp variation in both the neutral and positive scores. Evaluation of what method provided accurate result depends on several factors and is beyond the scope of this work.

*B. Text Mining Sentiment Analysis of Twitter data*

To demonstrate how the framework can be applied to model a machine learning classifier for predicting the sentiment of a given text, we collected about 11,431 tweets with the following keywords: medicine; prescription; over the counter; side-effects; online pharmacy; and antibiotics. The tweets were first cleaned then converted to a single corpus, with each tweet represented as a single document. Thus we have 11,431 documents making up the corpus. The corpus is then represented as sparse matrix for frequent term analysis and lexicon generation as described in part A.

We also performed lexicon sentiment analysis over the entire corpus as in part A, such that the generated sentiment scores containing 11,431 tweets classified with corresponding scores as described in TABLE 4.5 will serve as an initial datataset; for simplicity, we categorized scores greater than zero as positives and scores lower than zero as negatives.

TABLE 4.5. RANDOM PREVIEW OF POLARITY SCORES

| Score | Tweet |
|---|---|
| -1 | To everybody that knows me I think it's time for me to sign paperwork to take me off my antibiotics which will allow me to die in my home |
| 2 | Pretty sure we headed a hospital visit off today, fingers crossed these antibiotics help ASAP. |
| 0 | Took my antibiotics without eating and i was almost sick in work |
| -2 | Its ridiculas how many times i get ill, iv either got a cold on antibiotics , my emune system is rubbish |

The dataset was split into 75% i.e. 8573 labeled tweets as a training set and 25% i.e. 2858 labeled tweets as test set so that the classifier can be evaluated on data it had not seen previously. We apply the Naive Bayes algorithms as our baseline classifier. TABLE 4.6 is the representation of the result as a confusion matrix.

TABLE 4.6. NAÏVE BAYES CLASSIFIER RESULT

| Predicted | Actual | | Row Total |
|---|---|---|---|
| | Negative | Positive | |
| Negative | 531<br>0.639 | 180<br>0.089 | 711 |
| Positive | 300<br>0.361 | 1847<br>0.911 | 2147 |
| Column Total | 831<br>0.291 | 2027<br>0.709 | 2858 |

Looking at the table, we can see that 300 out of 831 positive messages (36 percent) were incorrectly classified as negative, while 180 of 2027 negative messages (8.9 percent) were incorrectly classified as positive; a total accuracy of about 83%. This performance will be used as a baseline for assessing other classifiers.

## Conclusion

We have demonstrated how machine learning techniques can be used to infer sentiments over social media data suggesting the views and experiences of drug and cosmetic product users. First a framework is developed for harnessing and tracking these views and experiences using text mining and sentiment analysis, we then conducted two case studies using the framework for comparison of sentiment analysis over three cosmetic brands coded as: Brand X, Brand Y and Brand Z and also over three products: Soap, Cream and Deodorant. A Naïve Bayes classifier was used to obtain a baseline result for assessing other classifiers. This paper reports a work in progress and the initial brand and product comparison results signify the usefulness of text mining and sentiment analysis on social media data while the use of machine learning classifiers for predicting the sentiment orientation provide a useful tool for users, product manufacturers, regulatory and enforcement agencies to monitor brand or product sentiment trends in order to act in the event of sudden or significant rise in negative sentiments.

Future work will consider comment spamming, comparing different machine learning sentiment classification performances, temporal analysis for detecting up or down trend of sentiment of a particular brand or product as well as clustering tweet and user sentiments by location.

## Acknowledgment

Special thanks to the Commonwealth Scholarship Commission through the Association of Commonwealth Universities (ACU) for providing studentship to the main author.